\documentclass[useAMS,twocolumn,trackchanges]{aastex63}

\usepackage{amsmath,amssymb}
\usepackage{natbib}
\usepackage{epsfig}
\usepackage{times}
\usepackage{rotate}
\usepackage{hyperref}
\usepackage{url}
\usepackage{subfigure}
\usepackage{cprotect}
\usepackage{xspace}
\usepackage{enumerate}
\usepackage{fancyhdr}
\usepackage{appendix}
\usepackage[pass]{geometry}
\usepackage{bm}
\usepackage{lineno}

\newcommand{\simgt}{\lower.5ex\hbox{$\; \buildrel > \over \sim \;$}}
\newcommand{\simlt}{\lower.5ex\hbox{$\; \buildrel < \over \sim \;$}}

\newcommand{\m}{${\rm M}_{\rm 200b}$ }

\received{--}
\revised{--}
\accepted{--}
\submitjournal{ApJ}

\begin{document}


\title{POLARIZATION STUDY OF GAMMA-RAY BINARY SYSTEMS}




    \author{HU XINGXING}
\email{huxx09791@hust.edu.cn}
\affiliation{Department of Astronomy, Huazhong University of Science and Technology}
    \author{TAKATA JUMPEI}
\email{takata@hust.edu.cn}
\affiliation{Department of Astronomy, Huazhong University of Science and Technology}



\begin{abstract}

The polarization of X-ray emission is a unique tool used to investigate the magnetic field structure around astrophysical objects. In this paper, we study the linear polarization of X-ray emissions from gamma-ray binary systems based on pulsar scenarios. We discuss synchrotron emissions from pulsar wind particles accelerated by a standing shock. We explore three kinds of axisymmetric magnetic field structures: (i) toroidal magnetic fields, (ii) poloidal magnetic fields, and (iii) tangled magnetic fields. Because of the axisymmetric structure, the polarization angle of integrated emission is oriented along or perpendicular to the shock-cone axis projected on the sky and swings by 360$^{\circ}$ in one orbit. For the toroidal case, the polarization angle is always directed along the shock cone axis and smoothly changes along the orbital phase. For the poloidal/tangled magnetic field, the direction of the polarization angle depends on the system parameters and orbital phase. In one orbit, the polarization degree for the toroidal case can reach the maximum value of the synchrotron radiation ($\sim70\%$), while the maximum polarization degree for poloidal/tangled field cases is several 10\%. We apply our model to bright gamma-ray binary LS 5039 and make predictions for future observations. With the expected sensitivity of the Imaging X-ray Polarimetry Explorer, linear polarization can be detected by an observation of several days if the magnetic field is dominated by the toroidal magnetic field. If the magnetic field is dominated by the poloidal/tangled field, significant detection is expected with an observation longer than 10 days.

\end{abstract}

\keywords{shock waves, gamma-rays, massive stars, mass-loss, neutron star}

\section{Introduction}
Gamma-ray binaries are binary systems composed of a compact object (black hole or neutron star) and a nondegenerate stellar companion, which in the cases studied here will be a high-mass star. The unique property of a gamma-ray binary system is nonthermal emission from the radio to very high-energy TeV bands, and its peak in the spectral energy distribution seems to be around MeV energy bands. Approximately 10 sources have been discovered, namely, PSR B1259-63/LS~2883 \citep{1992ApJ...387L..37J,1994ApJ...427..978C,2005A&A...442....1A},
PSR J2032+4127/MT91~213 \citep{2015MNRAS.451..581L,2017MNRAS.464.1211H,2017ApJ...836..241T,2018ApJ...867L..19A}, LS~5039 \citep{2012A&A...548A.103M,2009ApJ...697..592T,2006A&A...460..743A,2009ApJ...706L..56A},
LS~I+61$^{\circ}$303 \citep{2006Sci...312.1771A,2013ApJ...779...88A}, 1FGL~J1018.6-5856 \citep{2012A&A...541A...5H,2015ApJ...806..166A}, H.E.S.S. J0632+057 \citep{2009ApJ...690L.101H,2017ApJ...846..169L,2018PASJ...70...61M},
LMC~P3 \citep{2016ApJ...829..105C}, 4FGL~J1405.4-6119 \citep{2019ApJ...884...93C} and HESS J1832-093 \citep{2020arXiv200102701M,2020JHEAp..26...45T}. Two of them host young radio pulsars (PSRs B1259-63 and J2032+4127), but the compact objects of the other systems have not yet been identified (\citealt{2020PhRvL.125k1103Y,2021ApJ...915...61V} for LS~5039).

Two competing models have been proposed to explain the nonthermal nature of gamma-ray binaries, namely, the micro-quasar scenario or compact pulsar wind scenario. In the micro-quasar scenario, it has been discussed that a non-relativistic jet \citep[e.g.,][]{2015ApJ...801...55Y, 2016MNRAS.456.3638Y} or a relativistic jet \citep[e.g.,][]{2018MNRAS.481.1455K,2019A&A...629A.129M} is launched from the compact object. The interaction of the jet with a strong stellar wind \citep[e.g.,][]{2010A&A...512L...4P,2016MNRAS.456.3638Y, 2019A&A...629A.129M} or with clumpy stellar wind \citep[e.g.,][]{2009A&A...503..673A,2017A&A...604A..39D} forms a shock, where a part of the kinetic energy of the jet is converted into the internal energy of the jet fluid. The shock accelerates the leptons (electrons and probably positrons) to very high energy, and the accelerated particles produce nonthermal emission via the synchrotron radiation and inverse-Compton process \citep{2006A&A...451..259P,2010MNRAS.404L..55D,2018MNRAS.481.1455K}.

In the pulsar scenario \citep{1997ApJ...477..439T}, a relativistic pulsar wind, which is mainly composed of electron/positron pairs and a magnetic field, interacts with the wind/disk of the companion star. The interaction creates a standing shock between two winds, and the shock acceleration process produces a population of very high-energy electrons and positrons. Both scenarios assume synchrotron radiation and the inverse-Compton scattering process to explain the observed X-ray and  TeV emissions, respectively.

The radioemissions of the gamma-ray binary systems have been observed with an elongated morphology around the systems.
The position angle of the elongated morphology on the observer's sky changes with the orbital phase, and the change is repeatable by every orbit \citep{2006smqw.confE..52D,2008A&A...481...17R,2012IJMPS...8..372M,2017AIPC.1792d0018M}.
This evolution of the radiomorphology is consistent with the pulsar scenario, which predicts that
the position angle of the flow direction projected on the observer's sky changes by $360^{\circ}$ over the orbital phase \citep{2006sf2a.conf..133D,2013A&ARv..21...64D}. Evidence of a quasiperiodic oscillation in radio and X-ray emission, on the other hand, has been reported for the gamma-ray binary LS~I+61$^{\circ}$303 and may suggest a microquasar scenario for this system \citep{1997A&A...328..283P,2018MNRAS.476.2516N,2017MNRAS.471L.110J,2021arXiv210709610J}.

In addition to the spectrum and the light curve, the information of linear polarization of the nonthermal emission may provide a useful tool to discriminate between the models.
In particular, ongoing projects in X-ray and soft gamma-ray bands \citep{2016SPIE.9905E..1QZ,2016SPIE.9905E..17W,2019BAAS...51g.245M} will
provide a unique opportunity to probe the magnetic field structure of the emission region. In the pulsar model, the shocked pulsar wind forms a cone-like morphology around the pulsar \citep{2012AIPC.1505..398Z}, and its axis is perpendicular to the orbital axis,
which is defined as the direction normal to the orbital plane. On the observed sky-plane,
the direction of the shock-cone axis rotates 360$^{\circ}$ with the orbital motion of the pulsar. We will show in this paper that the polarization angle (hereafter P.A.) of the linear polarization for the synchrotron emission from the shocked pulsar wind will swing by 360$^{\circ}$ over one orbital revolution of the pulsar.

In the microquasar scenario, the evolution of the polarization along the binary orbit will depend on bending or precession of the jets. It has been discussed that the jet could be bent by the strong stellar wind \citep{2010A&A...512L...4P,2015ApJ...801...55Y,2019A&A...629A.129M}. Since the asymptotic jet direction will change with orbital motion, it is expected that the polarization characteristic evolves along with the orbital phase. The amplitude of the swing of the linear polarization may provide information about the degree of jet bending and hence the power of the jet relative to the stellar wind.

In this paper, we will study the linear polarization of the synchrotron emission in X-ray/soft gamma-ray bands of gamma-ray binaries based on the pulsar scenario. We will apply our model of the polarization to the binary system LS~5039, in which the companion star is an O-type star. Since an O-type star does not form a stellar disk (decretion disk), the compact object interacts only with the stellar wind. We therefore expect that the shape of the pulsar wind shock (e.g., opening angle of the shock cone) is stable, which can simplify our discussion. Moreover, LS~5039 is a powerful gamma-ray binary system with a short orbital period of $\sim3.9$~days, which makes polarization observations more practical.

In Section 2, we describe our calculation method for the synchrotron emission from the shocked pulsar wind and illustrate the Lorentz transformation of the Stokes parameters. To calculate the polarization properties, we introduce three different kinds of magnetic field geometries of the shock cone: (i) a toroidal magnetic field around the shock-cone axis, (ii) a poloidal magnetic field along the shocked flow and (iii) a tangled magnetic field.
The shock geometry and property of the shocked pulsar wind are briefly discussed in this section. In Section 3, we discuss the general properties of the polarization characteristics with the assumed magnetic field geometry. In Section 4, we apply the model to the bright gamma-ray binary LS~5039. By fitting the observed X-ray light curve with the synchrotron emission model, we predict
the polarization characteristics for LS~5039. A brief discussion of future observations is presented in Section 5.

\section{Model calculation}
\subsection{Lorentz transformation of polarization directions}
A unique characteristic of the gamma-ray binary is the temporal variation of the observed flux level along the orbital phase.  Based on the pulsar scenario, previous studies explain orbital variation by introducing the Doppler effect due to the motion of subrelativisitic shocked pulsar wind \citep{2010A&A...516A..18D,2014ApJ...790...18T,2020arXiv200404337H}. Under the condition that the stellar wind power is stronger than the pulsar wind (section~\ref{structure}), the Doppler effect enhances the observed emission around the inferior conjunction (hereafter INFC) and suppresses the emission around the superior conjunction (hereafter SUPC). Therefore, we consider the transformation of the polarization direction between the flow rest frame and laboratory frame; we use nonprimed and primed symbols to represent the quantities in the laboratory frame and in the flow rest frame, respectively.

The unit vector of direction of the electric field of the electromagnetic wave ($\bm{e}$)
is transformed as (equation C3 in \citealt{2003ApJ...597..998L})
\begin{equation}
\bm{e}=\frac{\left[\bm{e}'-\frac{\Gamma_\beta}{\Gamma_\beta+1}(\bm{e}'\cdot\bm{\beta})\bm{\beta}-\bm{\beta}\times (\bm{n}'_o\times \bm{e}')\right]}{1+\bm{\beta}\cdot\bm{n}'_o},
\label{etrans}
\end{equation}
where $\bm{n}'_o$ is the unit vector for the propagation direction of the wave (observer direction), $\bm{\beta}$ is the flow velocity of the shocked pulsar wind divided by the speed of light and $\Gamma_\beta=1/\sqrt{1-\beta^2}$ is the Lorentz factor of the flow. The unit vector of the propagation direction is transformed as (equation (4) in \citealt{2003ApJ...597..998L})
\begin{equation}
 \bm{n}_o=\frac{\bm{n}'_o+\Gamma_\beta\bm{\beta}\left[\frac{\Gamma_\beta}{\Gamma_\beta+1}(\bm{n}'_o\cdot\beta)+1\right]}{\Gamma_\beta(1+\bm{n}'_o\cdot\bm{\beta})}
 \end{equation}
We express the unit vector for the direction of the observer as
\begin{equation}
\bm{n}_o=\sin\theta_o\cos\phi_o\bm{e}_x+\sin\theta_o\sin\phi_o\bm{e}_y
+\cos\theta_o\bm{e}_z
\label{nobs}
\end{equation}
where we take the axis of $\bm{e}_z$ in the direction of the orbit axis and the axes of
$\bm{e}_x$/$\bm{e}_y$ on the orbital plane (Figure~\ref{geometry}). In this study, we define the north direction as the orbital axis projected on the sky and measure the P.A. as a counterclockwise direction from the north direction.

\subsection{Shock geometry and magnetic field structure}
 \label{structure}

\begin{figure}
\includegraphics[scale=0.4]{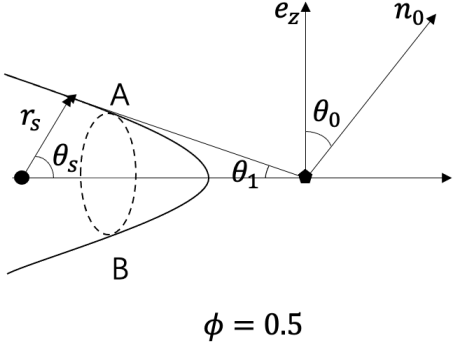}
\caption{Illustrative geometry of the shock-cone and binary system. The black filled circles and regular pentagon represent the pulsar and the companion star,
respectively. The vectors $\bm{e}_z$ and $\bm{n}_o$ represent the direction of the orbit axis, which is perpendicular to the orbital plane, and the observer, respectively. The solid curve represents the shock cone, and the distance from the pulsar to the shock, $r_s$, is described by equation~(\ref{rs1}) and equation~(\ref{rs2}). $\phi=0.5$ represents the orbital phase when the shock-cone axis is directed opposite to the direction of the observer (SUPC). The dashed circle represents the positions on the shock with a constant $\theta_{s}$.
The labels "A" and "B" represent the positions on the dashed circle and in the plane created
by the directions of the shock-cone axis and the line of sight. }
\label{geometry}
\end{figure}

In our calculation, we apply the shock-cone geometry discussed in \cite{1996ApJ...469..729C} and assume
that the shock cone is symmetric around the axis connecting the pulsar and companion star (Figure~\ref{geometry}). The distance to the shock measured from the pulsar is then calculated from

\begin{equation}
r_s(\theta_s)=D\sin\theta_s\rm{csc}(\theta+\theta_1),
\label{rs1}
\end{equation}
and
\begin{equation}
\theta_1\rm{cot}\theta_1=1+\eta(\theta_s\rm{cot}\theta_s-1),
\label{rs2}
\end{equation}
where $D$ is the separation between two stars and $\theta_s$ and $\theta_1$ are the angles to the emission region on the shock measured from the pulsar and companion star, respectively (Figure~\ref{geometry}). In addition, $\eta$ is the ratio of momentum rates of the pulsar wind and stellar wind. It has been estimated that the mass loss rate of an O-type star is on the order of $\dot{M}\sim 10^{-6}-10^{-7} M_{\odot}~\rm{year}^{-1}$ and that the typical terminal speed of the stellar wind is approximately 2000 km/s \citep{1996A&A...305..171P}. For a spin down power of $L_{sd}=10^{36-37}{\rm erg~s^{-1}}$, the momentum ratio is on the order of
\begin{eqnarray}
\eta&\equiv&\frac{L_{sd}}{\dot{M}v_w c}
\approx5.3\times 10^{-2}\left(\frac{v_w}{10^8{\rm cm~s^{-1}}}\right)^{-1} \nonumber\\
&&\left(\frac{L_{sd}}{10^{36}{\rm erg s^{-1}}}\right)\left(\frac{\dot{M}}{10^{-7}M_{\odot}{\rm year^{-1}}}\right)^{-1},
\label{ratio}
\end{eqnarray}
suggesting that the shock-cone confines the pulsar. We assume that the emission region
of the shocked flow is geometrically thin compared to the size of the binary separation, and therefore, we neglect the effect of the thickness of the shock region. We divide the shocked flow into many fluid elements and calculate the synchrotron emission for each fluid element.

A theoretical uncertainty in the current calculation is the structure of the magnetic field of the shocked pulsar wind flow. Since a random magnetic field structure expects unpolarized emission, we assume that the magnetic field averaged over the shock cone has a specific direction. It may be a reasonable assumption that the shock-cone axis is a geometrical axis, namely, the magnetic field is axisymmetric around the shock-cone axis. For example, the dynamics of the shocked pulsar wind with a toroidal magnetic field around the shock-cone axis were explored by \cite{2019MNRAS.490.3601B}. In our study, we explore  three simple situations;
\begin{enumerate}
\item a pure toroidal magnetic field around the shock-cone axis,
\item a pure poloidal magnetic field along the shocked flow, and
\item a tangled magnetic field, in which an angular distribution measured from a specific distribution is symmetric, and the magnitude is constant \citep[hereafter KS62]{1962SvA.....5..678K}.
\end{enumerate}
We also assume that the magnetic field structure is independent of the orbital phase.

\subsubsection{Stokes parameters for pure toroidal/poloidal magnetic field}
\label{uniform}
We express the unit vector of the magnetic field with $\bm{n}_b$ and the magnetic field component parallel to or perpendicular to the flow direction with $B_{||}$ and $B_{\perp}$, respectively. For a poloidal magnetic field structure, for example, we assume that
the magnetic field direction is parallel to the flow direction, namely, $B_{||}\neq0$ and $B_{\perp}=0$, or
$\bm{n}_b=\bm{n}_\beta$, where $\bm{n}_\beta$ represents the unit vector of the flow direction.
For a toroidal field structure, on the other hand, we assume that $B_{||}= 0$ and $B_{\perp}\neq0$ and choose
the direction of the magnetic field to be perpendicular to both the shock-cone axis and the flow direction. By
assuming the ideal MHD condition that there is no electric field in the flow rest frame, $\bm{E}'=0$, we
find the relation that $\bm{n}'_{b}=\bm{n}_{b}$ for the cases of poloidal or toroidal magnetic
field structure.

For each calculation point, we express the direction of the electric field vector of the synchrotron radiation as
\begin{equation}
\bm{e}'_i=\frac{\bm{n}'_o\times \bm{n}'_{b,i}}{|\bm{n}'_o\times \bm{n}'_{b,i}|}, 
\end{equation}
in the flow frame, where subscript $"i"$ represents the quantity of each calculation point. We then transform the direction of the polarization from the flow frame to the
laboratory frame using equation~(\ref{etrans}) and obtain the polarization direction in the laboratory frame.

To calculate the volume emissivity of the synchrotron emission in the flow rest frame, we assume a power-law distribution for the emitting electrons/positrons of the shocked pulsar wind,
\begin{equation}
N'(E')d\gamma'=KE'^{-p}d\gamma',
\label{dis}
\end{equation}
where $\gamma'$ represents the Lorentz factor of the particles in the flow rest frame.
The intensity of the synchrotron emission in the flow rest frame with a power-law distribution becomes
\begin{equation}
j'_{V,i}=\frac{1}{\Pi_0}\Phi(\nu')[B'\sin\mu']^{(p+1)/2}
\label{volum}
\end{equation}
where $\Pi_0=(p+1)/(p+7/3)$, $B'$ is the magnetic field strength and $\mu'$ is the angle between the observer and the magnetic field direction. In addition,
\begin{eqnarray}
\Phi(\nu')&=&\frac{\sqrt{3}2^{(p-1)/2}}{16\pi}
\Gamma\left(\frac{3p-1}{12}\right)\Gamma\left(\frac{3p+7}{12}\right)\nonumber \\
&&K\frac{e^3}{m_ec^2}\left(\frac{b}{\nu'}\right)^{(p-1)/2},
\end{eqnarray}
where $\nu'$ is the frequency of the electromagnetic wave and $b=3e/4\pi m_e^3c^5$. The Lorentz transformation of the volume emissivity and frequency are given by $j_{V,i}=\mathcal{D}^2j'_{V,i}$ and $\nu=\mathcal{D}\nu'$,
respectively, with the Doppler factor of
$\mathcal{D}=\Gamma_\beta^{-1}(1-\bm{n}_o\cdot\bm{\beta})^{-1}$. As a result,
the Stokes parameters at each calculation point in the laboratory frame is obtained from
\begin{eqnarray}
\left\{ \begin{array}{l}
\frac{dI_i}{ds}\\
\frac{dQ_i}{ds} \\
\frac{dU_i}{ds} \\
\end{array} \right\}
=\left\{ \begin{array}{l}
j_{V,i}\\
j_{V,i}\cos2\chi  \\
j_{V,i}\sin2\chi\\
\end{array} \right\}.
\label{stok}
\end{eqnarray}
For each orbital phase, we integrate equation~(\ref{stok}) over the volume of the emission region to obtain the total Stokes parameters, $I$, $Q$ and $U$. The polarization degree $\Pi$ (hereafter P.D. ), and P.A. $\chi$, for each orbital phase are calculated from
\begin{equation}
\Pi=\frac{\sqrt{Q^2+U^2}}{I}~~{\rm and}~~\tan2\chi=\frac{U}{Q},
\label{defp}
 \end{equation}
respectively.
\subsubsection{Tangled magnetic field}
\label{tangle1}
For the calculation of the polarization of the synchrotron emission from a tangled magnetic field, we refer to the study done by KS62. In this model, we assume that the magnetic field structure in each fluid element is symmetric around a specific direction, which is denoted as $\bm{h}'$, with an angular distribution $W(\theta')$, where $\theta'$ is the polar angle measured from the symmetry axis $\bm{h}'$. The volume emissivity of the synchrotron radiation after integrating over the contribution of the magnetic field with the different $\theta'$
is then calculated from
\begin{equation}
 j'_{V,i}=\frac{1}{\Pi_0}\Phi({\nu'})\int[B'(\theta')\sin\mu']^{(p+1)/2}W(\theta')d\Omega',
\end{equation}
where $d\Omega'$ represents a solid angle.

Since the magnetic field in the shocked pulsar wind may be frozen into the
plasmas, it may be reasonable to assume that the symmetric axis of local
magnetic field structure is parallel to the flow direction, $\bm{h}_i'\propto \bm{n'_{\beta,i}}$. Thus, the global magnetic field
is still axisymmetric to the shock-cone axis. We introduce the magnetic field anisotropy parameter, $\kappa$, which is defined by the ratio of the average magnetic field energy density of the two components,
\begin{equation}
\kappa\equiv \frac{<B'^2_{||}>}{<B'^2_{\perp}>/2},
\end{equation}
where
\[
<B'^2_{||}>=\int B'^2\cos^2\theta'W(\theta')d\Omega'
\]
and
\[
<B'^2_{\perp}>=\int B'^2\sin^2\theta'W(\theta')d\Omega',
\]
respectively. The factor of 2 in the denominator of equation (14) corresponds to two components on the perpendicular plane. When $\kappa<1$, the magnetic field is dominated by the component in
the plane normal to the symmetry axis. When $\kappa>1$, on the other hand, the magnetic field is dominated by the component parallel to the symmetry axis. For $\kappa=1$, two components share the energy of the magnetic field,
and the integrated emission over the solid angle has no linear polarization.

The magnetic field strength $B'(\theta')$ and the angular distribution $W(\theta')$ depend on anisotropy models; for example, KS62 assumes that the magnetic field $B'(\theta')$ is constant in magnitude, and they represent an angular distribution as a series in spherical harmonics,
\begin{equation}
W(\theta')=\frac{1}{4\pi}\left[1+3\sigma\cos\theta'+\frac{5}{2}\epsilon (3\cos^2\theta'-1)\right],
\label{ksdis}
\end{equation}
where factor $\sigma$ is related to the average field strength of the considered region and factor $\epsilon$ is related to the degree of anisotropy. \cite{1999ApJ...524L..43S}, who studied the linear polarization for the afterglow of the gamma-ray burst,
assumed the anisotropy of the magnetic field with
\begin{equation}
B'(\theta')\propto (\zeta^2\sin^2\theta'+\cos^2\theta')^{-1/2},~{\rm and}~W(\theta')\propto B'^3(\theta'),
\label{sari}
\end{equation}
where the parameter $\zeta$ determines the degree of anisotropy. We can find that the field anisotropy parameter, $\kappa$, is expressed as
\begin{equation}
\kappa=\frac{1+2\epsilon}{1-\epsilon}
\label{kaks}
\end{equation}
for KS62, and
\begin{equation}
\kappa=\zeta^2
\end{equation}
for  \cite{1999ApJ...524L..43S}.

\begin{figure}
  \includegraphics[scale=0.4]{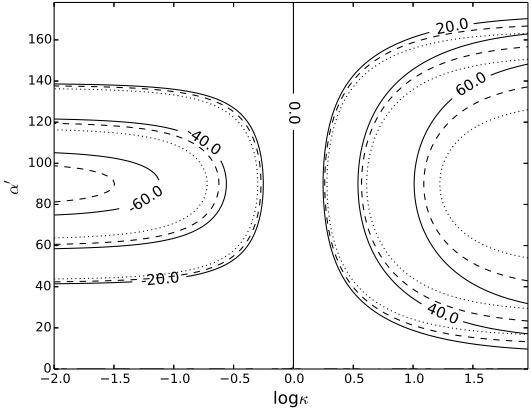}
\caption{ P.D. calculated with the anisotropy model of KS62 [entangled magnetic field described by equation~(\ref{linepl})] for a fluid element.
The horizontal and vertical axes represent the anisotropy and the angle of the observer measured from the symmetry axis $\bm{h}'$, respectively. The positive and negative P.D. correspond to the electric field vectors $\bm{e}'\propto \bm{n}'_o\times \bm{h}'$ and $\bm{e}'\propto \bm{n}'_o\times (\bm{n}'_o\times \bm{h}')$, respectively. The solid, dashed and dotted lines are the results for $p=2.5$, 2 and 1.5, respectively.}
  \label{fig:ks}
\end{figure}

With the anisotropy model of KS62, the volume emissivity integrated over the magnetic field distribution at each point [equation~(\ref{volum})] becomes
\begin{equation}
j'_{V,i}=j'_{o,i}\left[1+\frac{5}{4}\frac{p+1}{p+7}\epsilon(1-3\cos^2\alpha'_i)\right],
\label{ksvol}
\end{equation}
where $\alpha'$ is the angle between the line of sight and $\bm{h}'$, the symmetry axis of the anisotropic field, and
\[
j'_{o,i}=\frac{\Gamma\left(\frac{p+5}{4}\right)}
{\Gamma\left(\frac{p+7}{4}\right)}B_i'^{(p+1)/2}\frac{\sqrt{\pi}}{2}\frac{\Phi(\nu')}{\Pi_0}.
\]
The polarization degree is then
\begin{equation}
\Pi_i=\Pi_0\frac{15}{8}\frac{p+5}{p+7}\frac{\epsilon \sin^2\alpha'_i}{[1+\frac{5}{4}\frac{p+1}{p+7}
\epsilon(1-3\cos^2\alpha'_i)]}.
\label{linepl}
\end{equation}

For the P.A., it can be shown that when the magnetic field is mainly dominated by the component
in the plane normal to the symmetry axis ($\kappa<1$), the integrated emission at each point is polarized in a plane
that is normal to the line of sight and the symmetry axis $\bm{h}'$, namely,
$\bm{e}'\propto \bm{n}'_o\times \bm{h}'$.
When $\kappa>1$, on the other hand, the emission is polarized to the direction of the symmetry axis projected on
the observer sky, which can be expressed by $\bm{e}'\propto \bm{n}'_o\times (\bm{n}'_o\times \bm{h}')$.
The direction of the polarization in the laboratory frame is obtained using the transformation of
equation~(\ref{etrans}).

Figure~\ref{fig:ks} presents the P.D. calculated with the anisotropy model by KS62 [equation~(\ref{linepl})] for a calculation point and shows the dependency on the anisotropy parameter ($\kappa$) and the viewing angle ($\alpha'$) measured from the symmetry axis $\bm{h}'$. The solid, dashed and dotted lines show the results for the power-law indexes.
$p=2.5$, 2 and 1.5, respectively. The sign of the P.D. in the figure represents the directions of the electric field
vector. The P.D. becomes zero when $\kappa=0$ or when the observer is looking from the
symmetry axis $\alpha'=0^{\circ}$ or $180^{\circ}$.
With a fixed anisotropy parameter $\kappa$, the P.D. reaches the maximum value when the observer
angle from the symmetry axis is $\alpha'=90^{\circ}$, for which the anisotropy of the magnetic field measured by
the observer is the largest. We find in the figure that the predicted P.D.
{does not change much with the power-law index, $p$.}

We create the P.D. map on the $\kappa-\alpha'$ plane with the anisotropy model of \cite{1999ApJ...524L..43S} [equation~(\ref{sari})] and compare it with Figure~\ref{fig:ks} of KS62's model. We can see that the behavior of the P.D. on the $\kappa-\alpha'$ plane is similar to that of the KS62 model, except for an extreme case of $\kappa\gg1$. In the limit $\kappa\rightarrow \infty$,  using the magnetic field distribution [equation~(\ref{sari})] explored by \cite{1999ApJ...524L..43S}, we can see that the P.D. becomes $\Pi_0$. For KS62's anisotropy model represented by the function $W(\theta')$ [equation~(\ref{ksdis})], on the other hand, we can see in equation~(\ref{linepl}) that if the power-law index is not equal to $p=3$, the P.D. in the limit $\kappa\rightarrow \infty$ (or $\epsilon \rightarrow 1$) is smaller than $\Pi_0$. This indicates that the magnetic field component perpendicular to the symmetry axis always provides a nonnegligible contribution to the total emission for KS62's model. We also note that in the limit $\kappa\rightarrow 0$, the magnetic field is equally distributed into the plane perpendicular to $\bm{h}'$. With the function form of equation~(\ref{ksdis}), therefore, the P.D. of the local fluid element is smaller than $\Pi_0$. This is different from the poloidal or toroidal case
(section~\ref{uniform}), in which the local P.D. is equal to $\Pi_0$.

\subsection{Shocked pulsar wind}
\label{shock}
Detailed modeling of the spectra and orbital variations of gamma-ray binaries has been performed in previous studies \citep{2012ApJ...750...70T,2008AIPC.1085..253S,2007MNRAS.380..320K,2016MNRAS.463..495C,2017ApJ...838..145A}. In this paper, we simplify the dynamics of the shocked pulsar wind with a one-zone model and inject shock-accelerated particles
at the shock apex, $r_{s,0}$, for which $\theta	_1,\theta_s=0$ in equation~(\ref{rs1}).
We denote $L_{sd}$ as the spin-down power carried by the pulsar wind.
We assume that the pulsar wind particles before the shock are described by a monoenergetic distribution, and we calculate the proper density of the pulsar wind at the shock ($r=r_{s,0}$) from
\begin{equation}
   n_1=\frac{ L_{sd}}{4\pi u_1 \Gamma_1 r_{s,0} ^2 m_e c^3 (1+\sigma_w)},
 \end{equation}
where $\sigma_w$ is the ratio of the magnetic energy to the kinetic energy of the pulsar wind just before the shock, $\Gamma_1$ is the Lorentz factor of the bulk motion of the pulsar wind, and $u_1=\sqrt{\Gamma_1^2-1}$ is the radial four velocity. The magnetic field of the pulsar wind just before the shock is calculated from
\begin{equation}
   B_1=\sqrt {\frac{ L_{sd} \sigma_w}{r_{s,0}^2 c(1+\sigma_w)}}.
 \end{equation}
To evaluate the physical
quantities of the pulsar wind just after the shock,
we apply the results of the perpendicular MHD shock \citep{1984ApJ...283..694K,1984ApJ...283..710K} and assume a low $\sigma_w$ limit ($\sigma_w << 1$). The proper number density and the magnetic field strength just after the shock are expressed as
\begin{equation}
   n_2(r_s)=\frac{n_1 u_1}{u_2},
\label{n2}
 \end{equation}
and
\begin{equation}
   B_2(r_s)=3 B_1(1-4\sigma_w),
 \end{equation}
respectively, where $u_2$ is the four velocities of the shocked pulsar wind. We parameterize the speed of the shocked pulsar wind. Along the shocked flow, we assume $r B(r)=$constant and $r^2 n(r)=$constant.

Although we apply a one-zone model to obtain the magnetic field and number density in the downstream region, we consider a three-dimensional shock cone as the emission region. After we inject the fresh shocked particles at the shock-cone apex, we move the shocked flow along the 3-dimensional shock cone surface. The number density and magnetic field at the radial distance $r$ from the pulsar on the shock cone are obtained from the one-zone model described in the previous paragraph. We assume that the flow speed is subrelativistic ($\beta<0.5$), and we ignore the difference between the magnitudes of the magnetic field and number density in the laboratory frame and flow rest frame. In the rest frame, we assume that the particles are isotropic.

The energy distribution of the shocked particles in the downstream region will develop under the effect of the cooling processes of the inverse-Compton scattering process, synchrotron process and adiabatic expansion. We estimate that the Lorentz factor of the electrons/positrons that emit synchrotron photons with energy $E_X\sim 1$ keV is
\begin{equation}
   \gamma\sim 2.4\times 10^{5}\left(\frac{r}{r_{s,0}}\right)^{1/2}\left(\frac{B_2\sin\theta_{p}}{1{\rm G}}\right)^{1/2}
   \left(\frac{E_X}{1\rm{keV}}\right)^{1/2}.
 \end{equation}
The time scale of the synchrotron cooling of the electrons/positrons emitting X-rays becomes
\begin{equation}
   \tau_{syn}(r)=2160{\rm~s}\left(\frac{r}{r_{s0}}\right)^{3/2}\left(\frac{B_2\sin\theta_p}{1{\rm G}}\right)^{-3/2}
\left(\frac{E_X}{1\rm{keV}}\right)^{-1/2}
 \end{equation}
For the electrons/positrons emitting the 1-10~keV X-rays, we can see that the synchrotron cooling time scale is much longer than the dynamical time scale, which is $\sim r_{s,0}/(c/3)\sim 0.1{\rm AU}/(c/3)\sim 100$~s. This indicates that adiabatic cooling is more important than synchrotron cooling for X-ray emitting particles. However, the cooling process of inverse-Compton scattering could be dominant for some gamma-ray binary systems, and a single power law distribution could not describe the energy distribution of the X-ray emitting particles in the downstream region \citep{2014ApJ...790...18T}. In our current calculation, however, we only take the adiabatic cooling process into consideration and assume that the total particle energy is proportional to $r^{-2/3}$ with a constant power-law index for simplicity.
Although a more detailed calculation with inverse-Compton cooling is necessary, we expect that the result is not affected by the detailed cooling process since the predicted polarization characteristic weakly depends on the power-law index of the particle distribution, as Figure 2 indicates.

\section{Results}
\label{results}
As we discussed in Section~\ref{structure}, we assume that the magnetic field and flow of the shocked pulsar wind have an axisymmetric structure around the shock-cone axis. In our model, since the magnetic field is symmetric around the shock cone axis, the total emission integrated over the shock cone at each orbital phase is oriented along or perpendicular to that axis projected on the sky. Figure~\ref{geometry}, for example, represents the case of $\phi=0.5$. We note that $\phi=0$ and $\phi=0.5$ correspond to the INFC and SUPC, respectively. We define $\phi =\phi_t -\phi_o$, with $\phi_t$ being the true anomaly of the position of pulsar and $\phi_o$ being the azimuthal angle of the observer defined in equation~(\ref{nobs}). In this section, we assume a circular orbit, and therefore, the magnetic field strength at the shock region does not depend on the orbital phase.

\subsection{Toroidal magnetic field }
\label{toro}
\subsubsection{Polarization angle}
\label{toropa}
\begin{figure}
\centering
\includegraphics[scale=0.4]{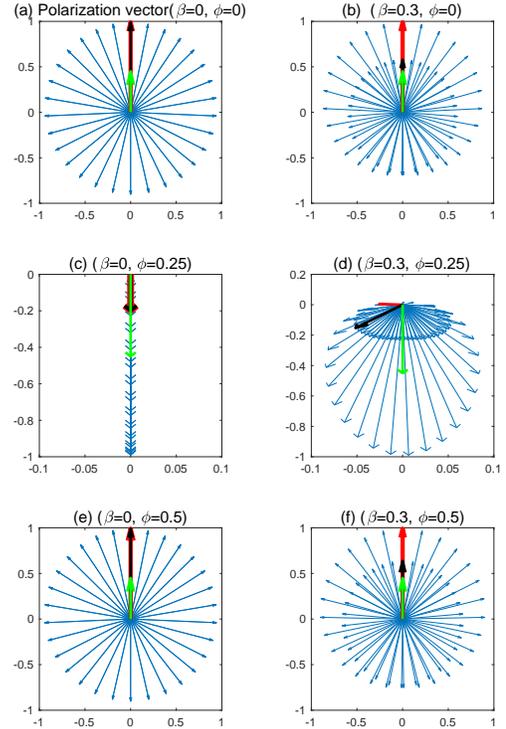}
\caption{
An illustrative example of the direction of the polarization vector (Q,U). Each blue arrow represents the direction of the polarization from a fluid element on the shock cone with a constant $\theta_s$ (e.g., dashed circle in Figure 1). The length of each arrow corresponds to $\sqrt{Q_i^2+U_i^2}$ from each emission point, and they are normalized by the maximum length.
The north direction is the projection of the orbital axis. If the polarization vector is along the north direction, P.A. equals to $0^{\circ}$. If the vector is heading south, P.A. equals to $90^{\circ}$.
The red and black colored arrows correspond to the emissions from the two points labeled by "A" and "B", respectively, in Figure~\ref{geometry}. The green arrow is the integrated polarization vector, and it only indicates the direction of the polarization vector. The results are for the toroidal magnetic field structure and at the orbital phases $\phi=0, 0.25$ and $0.5$ (Figure~\ref{geometry}). In addition, the left and right panels show the results for flow speeds of $\beta=0$ and 0.3, respectively. The inclination angle is $70^\circ$, and the angle between the flow direction and shock cone axis is $70^\circ$.}
\label{evec1}
\end{figure}

First, we discuss the toroidal magnetic field structure around the shock-cone axis. In such a case, we find that the total emission integrated over the shock cone is always polarized in the direction of the projection of the shock-cone axis on the sky.
By fixing the momentum ratio ($\eta$) and observer viewing angle ($\theta_o$), the polarization characteristic depends on the speed of the shocked pulsar wind since the Doppler effect changes the direction of the linear polarization and intensity at each emission point.
Figure~\ref{evec1} presents an illustrative example of the effects of the Doppler effect on linear polarization at each
emission point. Each blue arrow in the figure shows the polarization vectors (combination of $Q_i$ and $U_i$) of the synchrotron radiation from each calculation point (a fluid element) on a circular loop normal to the shock-cone axis and at the orbital phase of $\phi=0.5$ (for example, the dashed circle in Figure~\ref{geometry} represents such a circular loop). The length of each arrow represents $\sqrt{Q_i^2+U_i^2}$, which depends on the pitch angle $\mu'$ in equation (\ref{volum}) and the Doppler factor ($\mathcal{D}$); in each panel, the maximum length is normalized to unity. The vertical direction heading upwards in the figure corresponds to the direction of the orbital axis projected on the sky. The thick red and black arrows correspond to the emission from two points labeled by ``A'' and ``B'' in Figure~\ref{geometry}. The green arrows represent the direction of the polarization vector for the emissions integrated over the calculation points on the circular loop.

Figure~\ref{evec1}(e), for example, shows the case of $\beta=0$ and $\phi=0.5$. We can see that the length is the maximum at the red arrow (point ``A'') and black arrow (point ``B'') because the angle $\mu'$ between the observer and the direction of the toroidal magnetic field is the maximum at those two points. Since the linear polarization of the emission from A/B with the toroidal magnetic field directs toward the direction of the shock-cone axis projected on the sky, the integrated emission is also polarized in that direction. Although the distribution $\mu'$ on the circular loop changes for the different orbital phases, we find that the integrated emission over the calculation points with $\beta=0$ and the toroidal field is polarized to the shock-cone axis projected on the sky.

Figure~\ref{evec1}(f) shows the case of a flow speed of $\beta=0.3$.
In the figure, the two arrows that point in almost the same direction correspond to the emissions from symmetry two points relative to the shock-cone axis. Although the polarization vector (black arrow) of point B in Figure~\ref{geometry} is directed in the same direction as that (red arrow) of point A, the effect of Doppler boosting is stronger for the emission from point A than that from point B.
We find in the figure that the effect of Doppler boosting is the minimum for the emission from point A, and the polarization vector of the emission integrated over the calculation points is still directed to the direction of the shock-cone axis on the sky.
Comparing the panels of Figure~\ref{evec1}, we also find that for the different orbital phases, the polarization direction of the total emission with the toroidal magnetic field is not affected by the Doppler effect, and it is oriented to the direction of the shock-cone axis projected on the sky.

\begin{figure}
\includegraphics[scale=0.4]{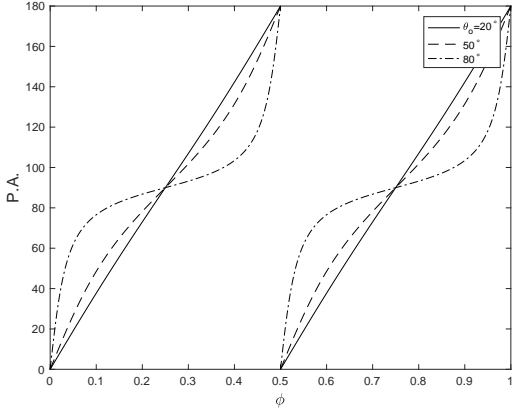}
\caption{Evolution of the P.A. over the orbital phase for the toroidal magnetic field structure.
The horizon axis represents the phase $\phi=\phi_t-\phi_o$. The solid, dashed and dotted lines are the results for inclination angles $\theta_o=20^{\circ}$, $50^{\circ}$ and $80^{\circ}$, respectively. The evolution is represented by equation~(\ref{para}).}
\label{pa}
\end{figure}

For the toroidal magnetic field, we can express the orbital evolution of the P.A. ($\chi$) measured from the orbital axis projected on the sky as
\begin{equation}
\cos\chi(\phi)=-\frac{\cos\theta_{o}\cos\phi}{\sqrt{1-\sin^2\theta_o\cos^2\phi}}.
\label{para}
\end{equation}
Figure~\ref{pa} presents the evolution of the P.A. for the different polar angles ($\theta_o$) of the observer. We see that the P.A. swings by $360^{\circ}$ over the orbital phase since the direction shock-cone axis projected on the sky does so. We note that since the P.A. in Figure~\ref{pa} is represented as a function of $\phi =\phi_t -\phi_o$, with $\phi_t$ being the true anomaly and $\phi_o$ being the azimuthal angle of the observer defined in equation~(\ref{nobs}),
the evolution of the P.A. with $\phi$ is independent of the eccentricity of the binary system. The P.A. approaches $90^{\circ}$ as the observer viewing angle approaches the edge-on view $\theta_o=90^{\circ}$. This can be understood because for the observer with a polar angle $\theta_o\sim 90^{\circ}$, the direction of the shock-cone axis on the sky is nearly perpendicular to the orbit axis, except for $\phi\sim 0$ or $\sim 0.5$.

\subsubsection{Polarization degree and intensity}

\begin{figure}
\includegraphics[scale=0.4]{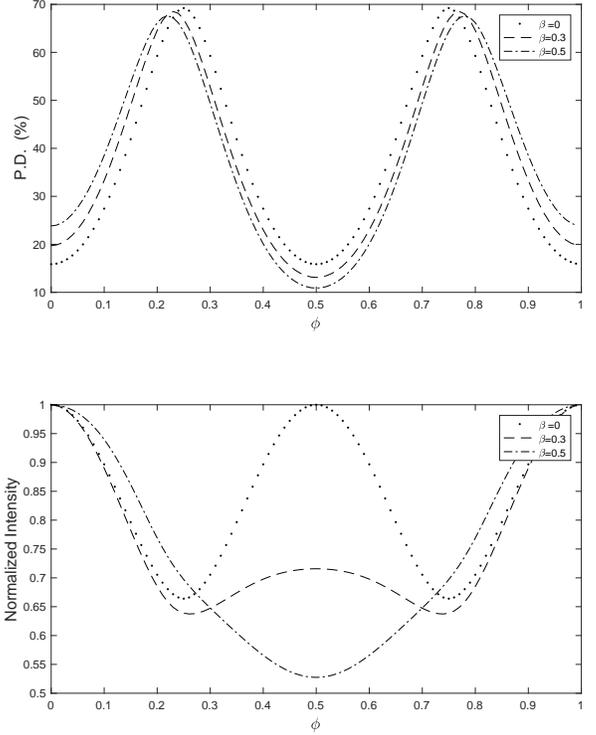}
\caption{Orbital variations of P.D. (top panel) and normalized intensity (bottom panel)
for a toroidal magnetic field structure. The dotted, dashed and dashed-dotted lines show the results for
$\beta=0$, 0.3 and 0.5, respectively. The system parameters for the results are the inclination angle of
$\theta_o=50^{\circ}$, the power-law index of $p=2$ and the momentum ratio of $\eta=0.1$}.
\label{syto}
\end{figure}

Within the framework of the current model, although the P.A. for the toroidal field structure does not depend on parameters of the
system (e.g., speed of the shocked pulsar wind, viewing geometry, etc.), the P.D. and the intensity depend on the parameters. Figure \ref{syto} shows the calculated orbital evolutions of P.D. (top panel) and intensity (bottom panel)
for a toroidal magnetic field with a flow speed of $\beta=0$ (dotted line), 0.3 (dashed line) and
0.5 (dashed-dotted line), respectively. As the top panel in the figure shows, the P.D. of the
toroidal magnetic field can reach a maximum value of $\Pi_0\sim 70\% $ for the power-law index $p=2$. The peak of the P.D. occurs at the orbital phase $\phi\sim 0.25$ and 0.75, where the direction of the shock-cone axis is perpendicular to the direction of the line of sight; hence, the direction of the magnetic field seen from the observer is a uniform structure. The Doppler effect with a finite speed of the flow slightly shifts the position of the maximum toward $\phi=0$ since it produces a dependency of the intensity on the emission point.

In the bottom panel of figure \ref{syto}, the orbital variation of the intensity for $\beta=0$ becomes the maximum at orbital phases $\phi=0$ and 0.5 since the average angle between the directions of the line of sight and
the magnetic field at each emission point (pitch angle that contributes to the ``observed'' emission) becomes the maximum.
 We note that in the
previous study of X-ray emission in the gamma-ray binary with the pulsar scenario, the orbital variation of the intensity is considered without the magnetic field structure \citep{2015A&A...581A..27D,2020A&A...641A..84M,2021A&A...646A..91H,2021A&A...649A..71H}, and thus a random magnetic field structure is implicitly assumed.
Since we assume that the shocked pulsar wind wraps the pulsar, the Doppler effect can enhance or suppress the observed emission at approximately $\phi=0$ (INFC) or $\phi=0.5$ (SUPC), respectively.

\subsection{Poloidal magnetic field}
\label{polo}
\subsubsection{Polarization angle}
\label{polo1}
\begin{figure}
\includegraphics[scale=0.4]{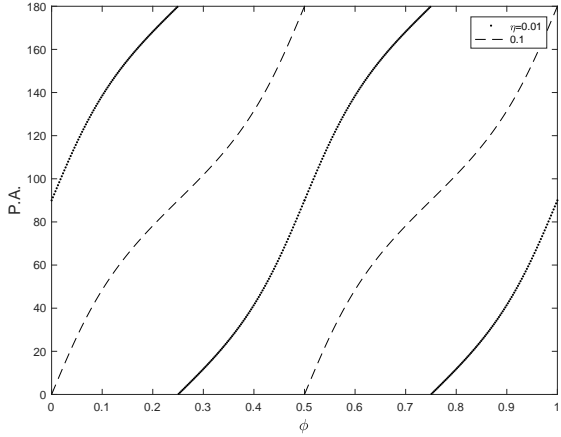}
\caption{Orbital variation of P.A. for $\eta=0.01$ (dotted line) and $\eta=0.1$ (dashed line)
with $\theta_o=50^{\circ}$. The magnetic field is parallel to the flow direction.
The solid line and dashed line are expressed by equations~(\ref{perp}) and ~(\ref{para}), respectively.
The results are for $\beta=0$. }
\label{eta}
\end{figure}

\begin{figure}
\includegraphics[scale=0.5]{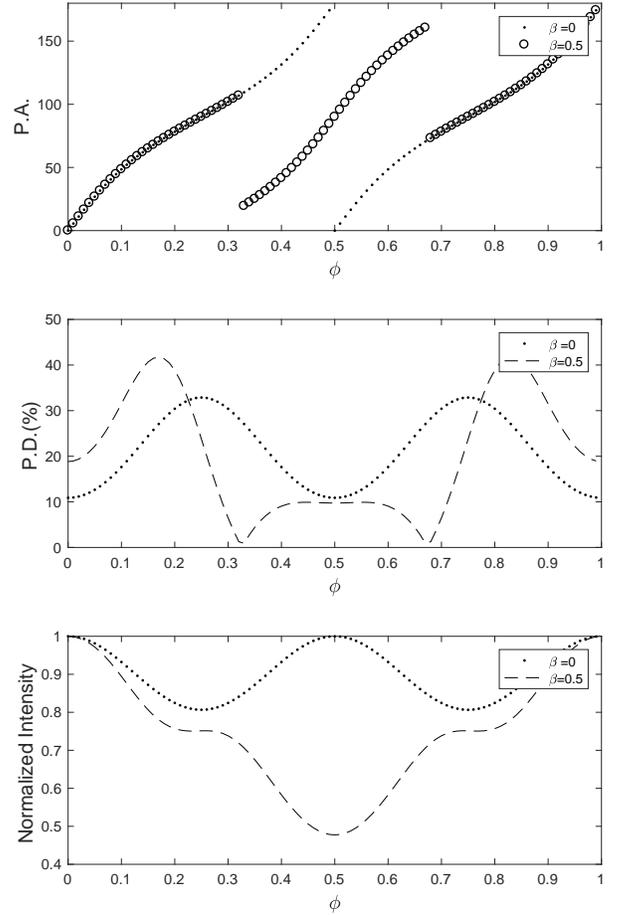}
\caption{
Example of the orbital variation for P.A. (top panel), P.D. (middle panel) and normalized intensity (bottom panel)
for a poloidal magnetic field structure. The flow speed is assumed to be $\beta=0$ for the dotted line and 0.5 for
open circles (top panel) or dashed lines (middle and bottom panels). The results are for a system inclination angle
of $\theta_o=50^{\circ}$ and a momentum ratio of $\eta=0.1$. }
\label{sypl}
\end{figure}

Another extreme situation is that the magnetic field is along the flow direction, and the projected magnetic field on the observer sky is still symmetric with respect to the direction of the projected shock-cone axis.
With the poloidal magnetic field structure, the electric field of the wave in the flow rest frame is oriented in the direction $\bm{e}'_i=(\bm{n}'_o\times \bm{n}_{\beta,i})/|\bm{n}'_o\times \bm{n}_{\beta,i}|$. Using equation~(\ref{etrans}) for the Lorentz transformation of the electric field vector, we can show that the electric field vector in the laboratory frame is expressed in the same form as it in the flow frame, namely,
\begin{equation}
\bm{e}_i=\frac{\bm{n}_0\times \bm{n}_{\beta,i}}{|\bm{n}_0\times \bm{n}_{\beta,i}|}.
\end{equation}

As we discussed in Section~\ref{toropa}, the P.A. for the toroidal field structure
does not depend on the system parameters and directs along the shock-cone axis projected on the sky. For the poloidal magnetic field, on the other hand, we can see that the polarization direction of the integration emission can be parallel to or normal to the shock-cone axis projected on the sky, and it depends on the opening angle of the shock cone and the flow speed. This is because the magnetic field can have two components parallel and perpendicular to the geometric axis (that is, the shock-cone axis). Figure~\ref{eta} shows the dependency of the P.A. evolution on the opening angle of the shock cone (or momentum ratio) by assuming $\beta=0$; the solid and dashed lines are the results for $\eta=0.01$ and $0.1$, respectively, with $\theta_o=50^{\circ}$.

For a smaller opening angle with momentum ratio $\eta<<1$, the shocked flow is more concentrated on the orbital plane, and the direction of the magnetic field projected on the observer sky is almost parallel to the shock-cone axis. In such a case, the integrated emission at each orbital phase is polarized to the direction perpendicular to the shock-cone axis projected on the sky, and the orbital evolution of the P.A. along the orbital phase (dotted line in Figure~\ref{eta} can be described by
\begin{equation}
\cos\chi(\phi)=-\frac{\sin\phi}{\sqrt{1-\sin^2\theta_o\cos^2\phi}}.
\label{perp}
\end{equation}
For a larger opening angle of the shock cone, on the other hand, the magnetic field perpendicular to the shock cone axis can be larger than the parallel component. In such a case, total
emission at each orbital phase can be oriented along the shock-cone axis on the sky, and the evolution of the P.A. along the orbital phase (dashed line in Figure~\ref{eta}) is described by equation~(\ref{para}).

Figure~\ref{sypl} summarizes the dependency of the polarization characteristics on the flow speed. In the figure, the top, middle and bottom panels show the orbital variation for the P.A., P.D. and intensity, respectively, and the dotted line and the dashed line are the results for flow speeds of $\beta=0$ and 0.5, respectively. For $\beta=0$, the total emission is always oriented in the direction along the shock-cone axis over the orbital phase. For $\beta=0.5$ (the open circle),
on the other hand, the P.A. changes the direction by 90$^{\circ}$ at
$\phi\sim 0.35/0.65$ and directs perpendicular to the shock-cone axis during $\phi=0.35-0.65$. We can see in the middle panel that the change in the P.A. occurs at the orbital phase,
at which total emissions are unpolarized ($P.D.=0$) with $Q=0$ and $U=0$.

\begin{figure}
\begin{center}
\includegraphics[scale=0.4]{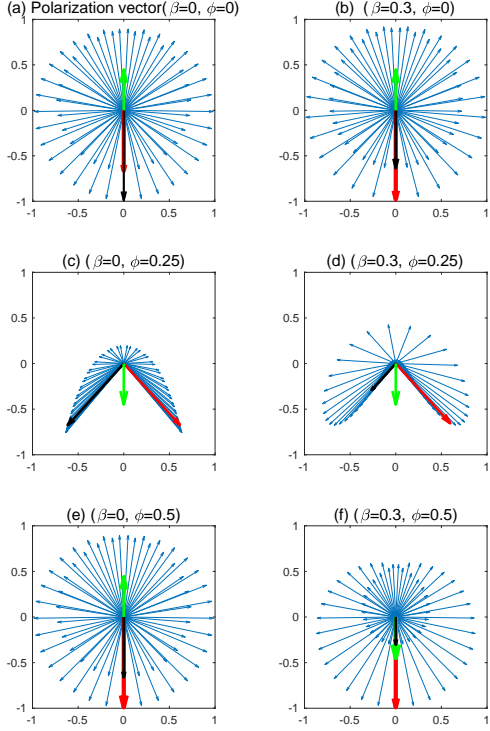}
\end{center}
\caption{Same as Figure~\ref{evec1}, but for the poloidal magnetic field, the magnetic field direction is along the flow direction. The green arrow in the right panel, which is downwards, indicates that the integrated emission is polarized perpendicular to the shock cone-axis projected on the sky.}
\label{evec}
\end{figure}

Similar to Figure~\ref{evec1} of the toroidal field, Figure~\ref{evec} represents the polarization vector on the sky for the poloidal magnetic field at orbital phases $\phi=$0 (top panel), 0.25 (middle panel) and $0.5$ (bottom panel). In the bottom panels, the red arrow and black arrow correspond to the polarization vectors from points "A" and "B", respectively, in Figure~\ref{geometry}. As Figure~\ref{evec}(e) shows, the lengths of polarization vectors of points "A" and "B are different because of different pitch angles $\mu '$.
The length of the vector is maximum at position A, where P.A. is perpendicular to the orbit axis, but the contribution from other positions makes the P.A. of the total emission parallel to projection of the orbit axis.

Compared with Figure~\ref{evec}(e), for $\beta=0.3$, $\phi=0.5$ in Figure~\ref{evec}(f), the Doppler boosting effect makes the difference between the length of polarization vectors from point "A" (red arrow) and point "B" (black arrow) greater.
The direction of the integrated polarization vector in Figure~\ref{evec} changes by $180^{\circ}$, indicating that the polarization direction changes $90^\circ$. However, for $\phi=0$ and $0.25$, the Doppler effect does not change the polarization direction. Unlike the case of the toroidal magnetic field, the Doppler effect affects the direction of the polarization for the case of the poloidal magnetic field.

\subsubsection{Polarization degree and intensity}
\label{pintensity}

By comparing the middle panel of Figure~\ref{sypl} with the top panel of Figure~\ref{syto}, we find that the P.D.
for a poloidal case with the maximum value of $\sim 20-30\%$ is smaller than
the case of the toroidal field. Because the projected magnetic field of the sky is more diverse than the toroidal magnetic field, the depolarization in the total emission is greater.
As we discussed in section~\ref{polo1}, the Doppler effect with $\beta=0.5$ (dashed line in the middle panel)
also causes depolarization of the emission at the orbital phase $\phi\sim 0.35-0.65$ and produces unpolarized emission at $\phi\sim 0.35$ and $0.65$, where the Stokes parameters $Q=0$ and $U=0$ occur simultaneously.

By comparing the light curve of the poloidal case (bottom panel of Figure~\ref{sypl}) with that of the toroidal case (Figure~\ref{syto}), we also find that the amplitude of the light curve of the poloidal field is smaller than
that of the toroidal case. This is because the projected poloidal
magnetic field on the sky
is more diverse than that case of the toroidal magnetic field, and therefore, the orbital variation of the average pitch angle that contributes to the observed emission is smaller. In short, the poloidal magnetic field predicts a smaller P.D. and a smaller amplitude of the light curve compared to those for the toroidal magnetic field.

\subsection{Tangled magnetic field}
\label{tangle}
In Sections 3.1 and 3.2, we discussed the polarization properties with two extreme cases for the toroidal field around the shock-cone axis and poloidal field along the flow direction. In this section, we discuss the polarization characteristic
with a tangled magnetic field structure with the KS62 model, which may provide a more realistic
situation. As described in Section~\ref{tangle1}, we assume that the symmetric axis of the tangled field is along
the flow direction, and we characterize the anisotropy
of the magnetic field structure with the parameter $\kappa$. When $\kappa<1$, the
configuration of the magnetic field is dominated by the component in the plane normal to
the local fluid velocity. For $\kappa>1$, on the other hand, the magnetic field is
dominated by a component parallel to the flow direction. An unpolarized emission is
expected for $\kappa=1$.

\subsubsection{Polarization angle}

\begin{figure}
 \includegraphics[scale=0.46]{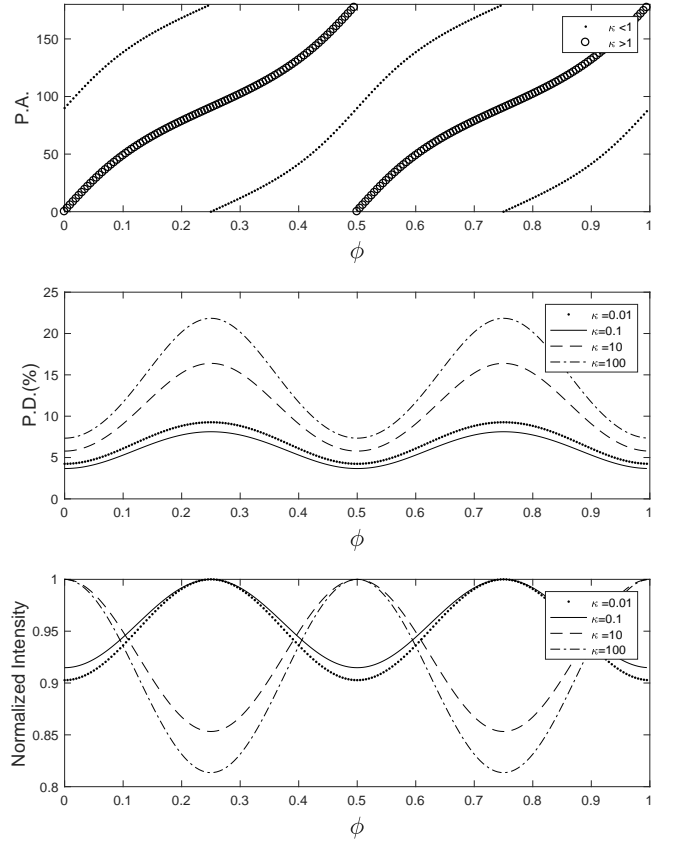}
\caption{Polarization characteristics for a tangled magnetic field structure, in which the symmetrical axis of the magnetic field is parallel to the flow direction. The results are for $\beta=0$, $\eta=0.1$ and $\theta_o=50^{\circ}$.}
\label{syks1}
\end{figure}

\begin{figure}
 \includegraphics[scale=0.46]{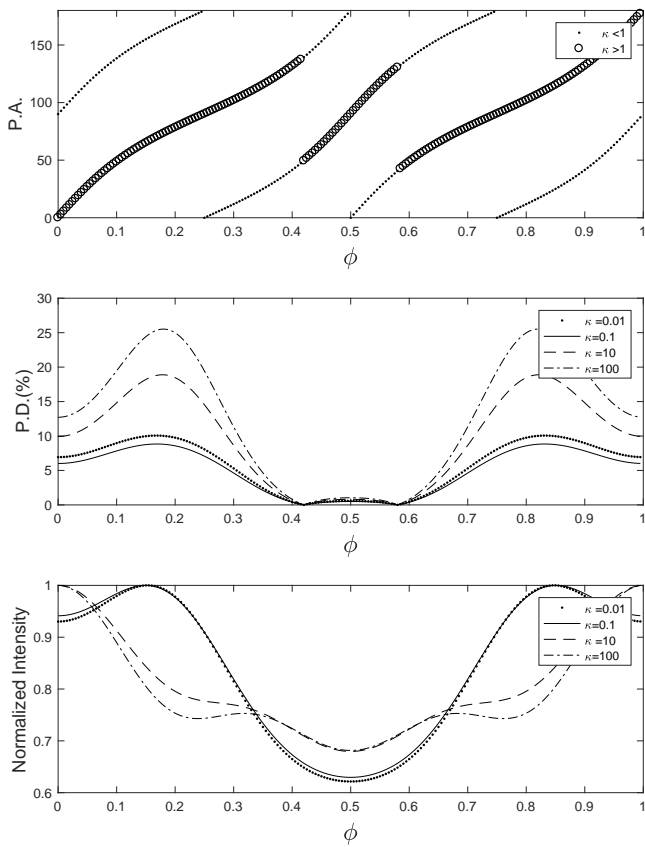}
\caption{Same as Figure~\ref{syks1}, but for $\beta=0.3$.}
\label{syks2}
\end{figure}
As described in Section~\ref{tangle1}, the P.A. of the integrated emission
over the magnetic field direction at each emission point is described
by $\bm{e}_i'\propto \bm{n}_o'\times \bm{h}'_i$ for $\kappa>1$ and by $\bm{e}'_i\propto  \bm{n}'_o\times ( \bm{n}_o'\times \bm{h}'_i)$ for $\kappa<1$. By assuming $\bm{h}_i'\propto \bm{n}_{\beta,i}$, the electric vector of electromagnetic waves produced by synchrotron emission for $\kappa>1$ becomes $\bm{e}_i'\propto \bm{n}_o'\times \bm{n}_{\beta,i}$, and its P.A. is represented by the case of the pure poloidal magnetic field discussed in Section~\ref{polo}. For the case $\kappa<1$, since the polarized
direction of the local emission is expressed by $\bm{e}_i'= \bm{n}'_o\times (\bm{n}'_o\times \bm{n}_{\beta,i}$), the P.A. is changed by 90$^{\circ}$ from the case of $\kappa>1$ (hence, we will omit the plots for the case of $\kappa<1$ in some figures).

Figures~\ref{syks1} and~\ref{syks2} summarize the dependency of the polarization characteristics on the anisotropy parameter, $\kappa$, with $\beta=0$ and 0.3, respectively, and the figures show the results for the momentum ratio of $\eta=0.1$ and the system inclination angle of $\theta_o=50^{\circ}$. With these system parameters of $\eta$ and $\theta_o$, total emission from the flow with $\beta=0$ (Figure~\ref{syks1}) is oriented
in the direction of the shock-cone axis for $\kappa>1$ (open circles in the top panel), which is consistent with the case for the pure poloidal magnetic field structure discussed in section~\ref{polo} and Figure~\ref{eta}. When $\kappa<1$, on the other hand, the polarized direction is perpendicular to the shock-cone axis, as the dotted line in the top panel of Figure~\ref{syks1} shows.

In Figure~\ref{syks2} for $\beta=0.3$, we see a change in polarization direction due to the Doppler effect, as discussed in Section~\ref{polo1}. We find that the orbital phase, where the change in the P.A. by 90 occurs, does not depend on the anisotropy parameter $\kappa$. This is because the Stokes parameters $Q_i$ and $U_i$ at each calculation point can be described by $Q_i=\epsilon (\kappa)C_{q,i}$ and $U_i=\epsilon(\kappa) C_{u.i}$, respectively, where proportional factors $C_{q,i}$ and $C_{u,i}$ are independent of the anisotropy parameter, as indicated by equation~(\ref{linepl}). If the anisotropy parameter $\kappa$ is constant over the emission region, then the total Stokes parameters are described by $Q=\epsilon \sum C_{q,i}$ and $U=\epsilon \sum C_{u,i}$, indicating that the orbital position of $Q=U=0$ is independent of the anisotropy parameter.

\subsubsection{Polarization degree and intensity}
The middle panels of Figures~\ref{syks1} and~\ref{syks2} show the calculated P.D. over the orbital phase for $\beta=0$ and 0.3, respectively. As we discussed in Section~\ref{tangle1}, the current model of the tangled magnetic field structure assumes that the magnetic field component perpendicular to the symmetry axis always provides a nonnegligible contribution to total emission, even for $\kappa\rightarrow \infty$. Hence the predicted P.D. of the tangled magnetic field
is smaller than that of the poloidal case. For example,
the peak of the P.D. for $\beta=0.3$ and $\kappa=100$ (dashed-dotted line in Figure~\ref{syks2})
is approximately 20\%, while that for the poloidal field (the dashed line in Figure~\ref{sypl}) is approximately $30\%$. In Figures~\ref{syks1} and~\ref{syks2}, we also find that when the magnetic field component perpendicular to the symmetry axis dominates, $\kappa\ll1$, the P.D. can reach only several~\% over the orbital phase.

The bottom panels in Figures~\ref{syks1} and~\ref{syks2} summarize the calculated light curves for the tangled magnetic field. With the zero flow speed $\beta=0$ of Figure~\ref{syks1}, we can find that the orbital variation shows a double peak structure, but the peak positions for the cases of $\kappa>1$ and $\kappa<1$ are shifted by the orbital phase $\delta \phi=0.5$. Without the Doppler effect,
the orbital modulation of the intensity is determined by the orbital variation of the angle $\alpha'$ in equation~(\ref{ksvol}), which is the inclination angle of the line of sight measured from
the symmetry axis $\bm{h}'$. Since we assume that the symmetry axis is parallel to the flow direction, the averaged inclination angle $\alpha'$ over the emission region becomes the maximum value at the orbital phases $\phi=0.25$ and $0.75$. For the $\kappa>1$ ($\epsilon>0$),
Since the magnetic field is dominated by the parallel component, the averaged pitch angle that contributes to the observed synchrotron emission reaches the maximum value at orbital phases $\phi=0.25$ and $0.75$. As a result, the calculated
light curve has a peak at the orbital phases $\phi=0.25$ and $0.75$. This orbital variation of the intensity is similar to the case of the poloidal case.

For $\kappa<1$ ($\epsilon<0$), on the other hand, the local magnetic field is dominated by the perpendicular component, and the averaged pitch angle is the minimum at the orbital phases $\phi=0.25$ and $0.75$ (and the maximum
at the orbital phase $\phi=0$ and $0.5$). This pattern of the orbital variation is similar to the case of the toroidal magnetic field (Figure~\ref{syto}), and two peaks appear at $\phi=0$ and $0.5$. For $\beta=0.3$ of Figure~\ref{syks2}, we can see that the Doppler effect enhances and suppresses the emission around the orbital phases $\phi=0$ and 0.5, respectively.

\section{Application to LS~5039}
\label{ls5039}
In this section, we apply our model to a bright gamma-ray binary system, LS~5039. LS~5039 is composed of an unknown compact object and an O-type main sequence star and has an orbital period of $P_{orb}\sim 3.9$~days. The emission from LS~5039
has been observed with an energy flux $F_X\sim 10^{-11}~{\rm erg~cm^{-2}s^{-1}}$ in the 1-10~keV bands~\citep{2009ApJ...697..592T} and $\sim 10^{-10}~{\rm erg~cm^{-2}s^{-1}}$ in the MeV bands~\citep{2014A&A...565A..38C}, which could enable a measurement of the polarization property with future observations. For the orbital phase of
LS~5039, the periastron is conventionally defined as the position of the zero orbital phase
($\Phi_{peri}=0$), and the INFC occurs at $\Phi_{INFC}\sim 0.6-0.8$. \cite{2005MNRAS.364..899C} measured the eccentricity as $e\sim 0.35\pm 0.04$, while \cite{2011MNRAS.411.1293S} provided
$e=0.24\pm 0.08$ through an optical observation. In this study, we fix the eccentricity and the position of the INFC at $e=0.25$ and $\Phi_{INFC}=0.7$, respectively. We note that $\phi=0$ and $\phi=0.5$ in the previous sections correspond to
$\Phi_{INFC}=0.7$ and $\Phi_{SUPC}\sim 0.05$, respectively.

\cite{2020PhRvL.125k1103Y} obtained evidence of a period of $\sim 9$~s and an increase in the period by a rate of $\sim 3\times 10^{-10}~\rm{s~s^{-1}}$ in SUZAKU/NuSTAR data. If these timing properties are true, the compact object is a magnetar-type neutron star with a spin-down power of $L_{sd}\sim 10^{34} {\rm erg~s^{-1}}$. It is, however, suggested that the significance of the signal is not high enough to claim it as a detection period of $\sim 9$~s \citep{2021ApJ...915...61V}. In this study, we apply a standard pulsar scenario of a gamma-ray binary system and assume that the location of the shock is determined by the pressure balance between the pulsar wind and the stellar wind. With the scenario of a rotation-powered pulsar, the spin-down power is expected to be $\sim 10^{36-37}\rm{erg~s^{-1}}$ to explain the observed luminosity of LS~5039 with a source distance of $d\sim 2.5$~kpc \citep{2007ApJ...671L.145S,2009ApJ...697..592T,2014ApJ...790...18T,2020arXiv201211578B}. Since the mass loss rate of an O-type star can reach approximately $10^{-6} M_{\odot}~{\rm year^{-1}}$ \citep{1996A&A...305..171P}, the momentum ratio defined by equation~(\ref{ratio}) can be on the order of $\eta=0.01$. In this study, therefore, we fix the momentum
ratio at $\eta=0.01$. We assume the speed of the shocked pulsar wind and Earth viewing angle
as $\beta=0.5$ and $\theta_o=70^{\circ}$, respectively, to reproduce the amplitude of
the X-ray light curve of LS~5039.

\begin{figure}
\includegraphics[scale=0.45]{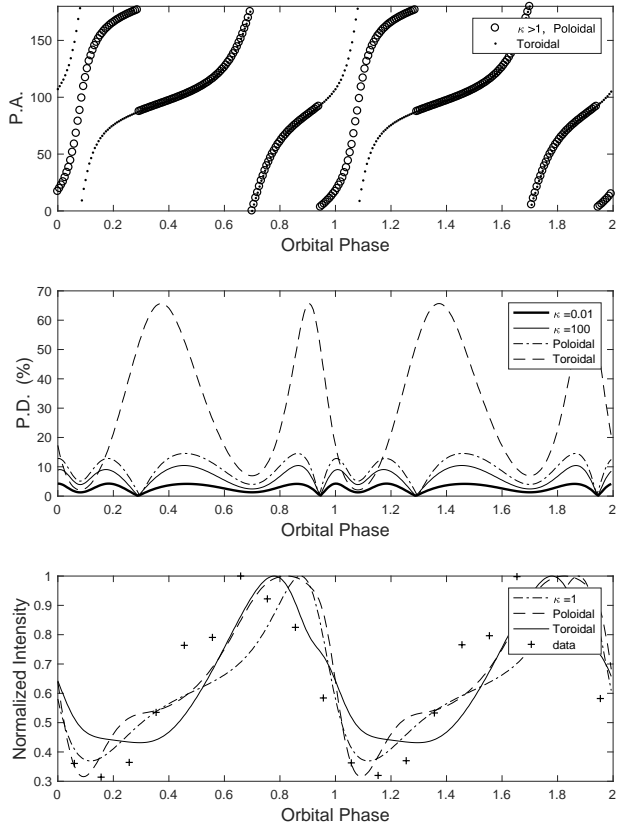}
\caption{Predicted polarization characteristic for LS~5039. Top: P.A. for a $\kappa>1$ and poloidal field structure (open circle symbol) and for a toroidal field structure (dotted symbol). Two symbols are overlapped between the orbital phase $\sim 0.6-0.8$. The P.A. for $\kappa<1$ is changed by $90^{\circ}$ from the P.A. for $\kappa>1$. Middle: P.D. for the different magnetic field structure. Bottom: Light curve of the X-ray emission from LS~5039.
The polarization characteristics for $\eta=0.1$ with $\theta_o=70^{\circ}$, $\beta=0.5$ and eccentricity are 0.25. Data of normalized intensity is cited from \cite{2009ApJ...697..592T}. The results are for $\theta_o=70^{\circ}$ and $\eta=0.01$. In addition, the periastron and INFC occur at the orbital phases of $\Phi_{peri}=0$ and $\Phi_{INFC}=0.7$, respectively. }
\label{symd}
\end{figure}

Figure \ref{symd} summarizes the predicted polarization characteristics for different structures of the magnetic field. In the current model, the light curve is affected by\
(i) the Doppler boosting effect, (ii) the magnetic field strength at the shock and (iii) the averaged pitch angle of the electrons that contribute to the observed synchrotron emission.
 The Doppler boosting effect tends to create a peak at the INFC in the simplified emission model \citep{2010A&A...516A..18D,2014IJMPS..2860169K,2014ApJ...790...18T}, while the effect of the magnetic field at the shock tends to enhance the emission at the periastron. In the bottom panel of Figure~\ref{symd}, for example, we can see that the intensity peak for a random magnetic field (dashed-dotted line for $\kappa=1$) appears to be the orbital phase between the INFC and SUPC, since we use the averaged value corresponding to $\sin ^2 \theta_p = 2/3$) for the pitch angle over the orbital phase.
We find that the light curves for $\kappa>>1$ and $\kappa<<1$ are similar to those of the poloidal magnetic field (dashed line in the figure) and toroidal magnetic field (solid line), respectively.

For the toroidal magnetic field structure, the light curve (solid line) shows a peak at the INFC. This is because the averaged pitch angle of the electrons that produces the observed emission obtains the maximum value at the INFC for the toroidal magnetic field, as discussed in Section~\ref{toro} (Figure~\ref{syto}).

In Figure~\ref{symd}, we see that although the calculated light curves for LS~5039 show a weak dependency on the magnetic field structure, the behavior of the predicted P.D. and P.A. are very different for different magnetic field structures. In the top panel, we summarize the predicted P.A. for a tangled field $\kappa>1$ and poloidal magnetic field (circle symbols) and for a toroidal magnetic field (dot symbols). We note that the P.A. for $\kappa<1$ is changed by $90^{\circ}$ for the case of $\kappa>1$. As we discussed in Section~\ref{toro}, the current model predicts that if the shock region is dominated by the toroidal magnetic field structure, the emission is polarized to the direction of the shock-cone axis projected on the sky, and its P.A. shows a smooth evolution over the orbital phase. For the tangled magnetic field ($\kappa>1$) and the poloidal magnetic field (circle symbol in the top panel), on the other hand, a jump in P.A. is expected, as discussed in Sections~\ref{polo} and~\ref{tangle}. With the system parameters ($\eta=0.01$, $\beta=0.5$ and $\theta_o=70^{\circ}$), the current model predicts that a jump in P.A. appears  at $\phi\sim 0.3$ and $0.9$. The P.A. evolution of two cases overlaps in $\phi \sim 0.3-0.9$, as shown in the top panel of Figure~\ref{symd}.

As we can see in the middle panel of Figure 11, the P.D. with a toroidal magnetic field can reach a maximum polarization degree $\Pi_0\sim 70\%$, while those of the tangled magnetic field and poloidal field are much smaller than the maximum value. Hence, a measurement of the emission with a larger P.D. by future observation would tell us that the magnetic field is dominated by the toroidal field component around the shock-cone axis. The model predicts that the peak of the P.D. for the toroidal magnetic field occurs at the orbital phase of $\Phi\sim 0.4$ and 0.9, where the direction of the shock-cone axis is perpendicular to the direction of the line of sight. For the tangled/poloidal field, on the other hand, the peak of P.D. is shifted toward the
superior conjunction ($\Phi_{SUPC}\sim 0.1$).

\begin{figure}
\includegraphics[scale=0.5]{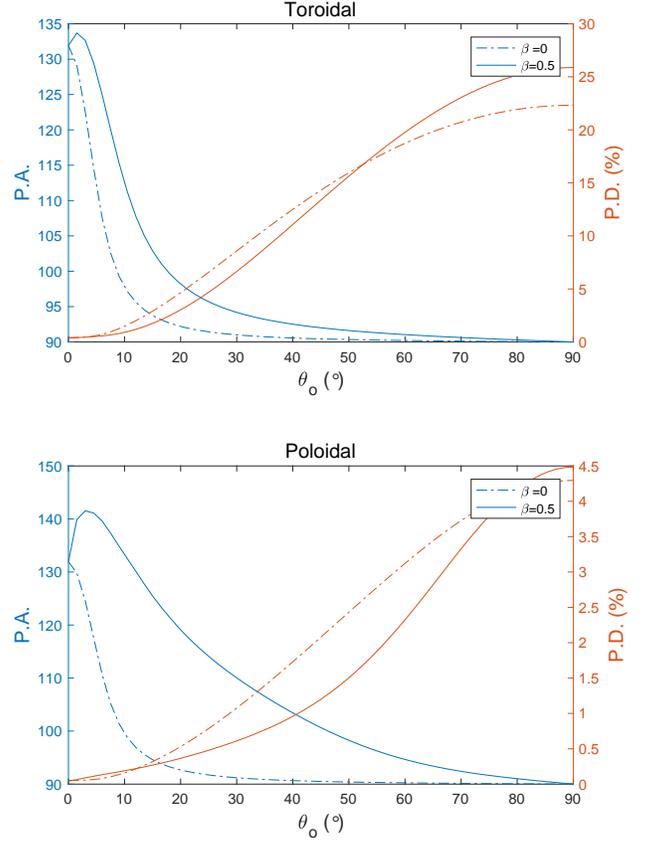}
\caption{Characteristics for orbital averaged P.D. and P.A. with the tangled magnetic field structure.
Top panel: Dependency of P.A. (blue color) and P.D. (red color) on the inclination angle $\theta_{o}$ for $\beta=0$(dash-dot line) and $\beta=0.5$(solid line) in the toroidal case. The results are for $\eta=0.01$.
Bottom panel: Dependency of P.A. (blue color) and P.D. (red color) on the anisotropy parameter $\kappa$ for $\beta=0$ (dash-dot line) and 0.5 (solid line). The results are for $\eta=0.01$. }
\label{sytp}
\end{figure}

Finally, we investigate the polarization characteristic averaged over the orbital phase with various system parameters. We find that the predicted polarization characteristic does not change much in the range of the eccentricity $0.2<e<0.35$, which is suggested for LS5039.
Figure~\ref{sytp} shows the dependency of P.A. (blue) and P.D. (red) inclination angle in the toroidal case (top panel) and poloidal case (bottom panel).
We can see in the figure that the dependencies of the P.A. on the inclination angle $\theta_{o}$ of the toroidal case and poloidal case are similar to each other. For the inclination angle $\theta_o\sim 0^{\circ}$, the P.A. of the integrated emission tends to be the direction of the semimajor axis or the orbit axis projected on the sky. With the parameters e=0.25 and $\Phi_{INFC}=0.7$, it is approximately $130^{\circ}$.
As we can see in the figure, P.D. increases with the inclination angle. This is expected because the polarization vectors for different orbital phases concentrate more in one direction for a larger inclination angle.
We find that the P.D. of the toroidal case can reach $>10 \%$ if $\theta_{o} > 30^{\circ}$, while that of the poloidal case is less than 10\%.
By comparing the results of $\beta=0$ (solid line) and $\beta=0.5$ (dashed line), the average polarization properties are not sensitive to the flow speed.

\begin{figure}
\includegraphics[scale=0.5]{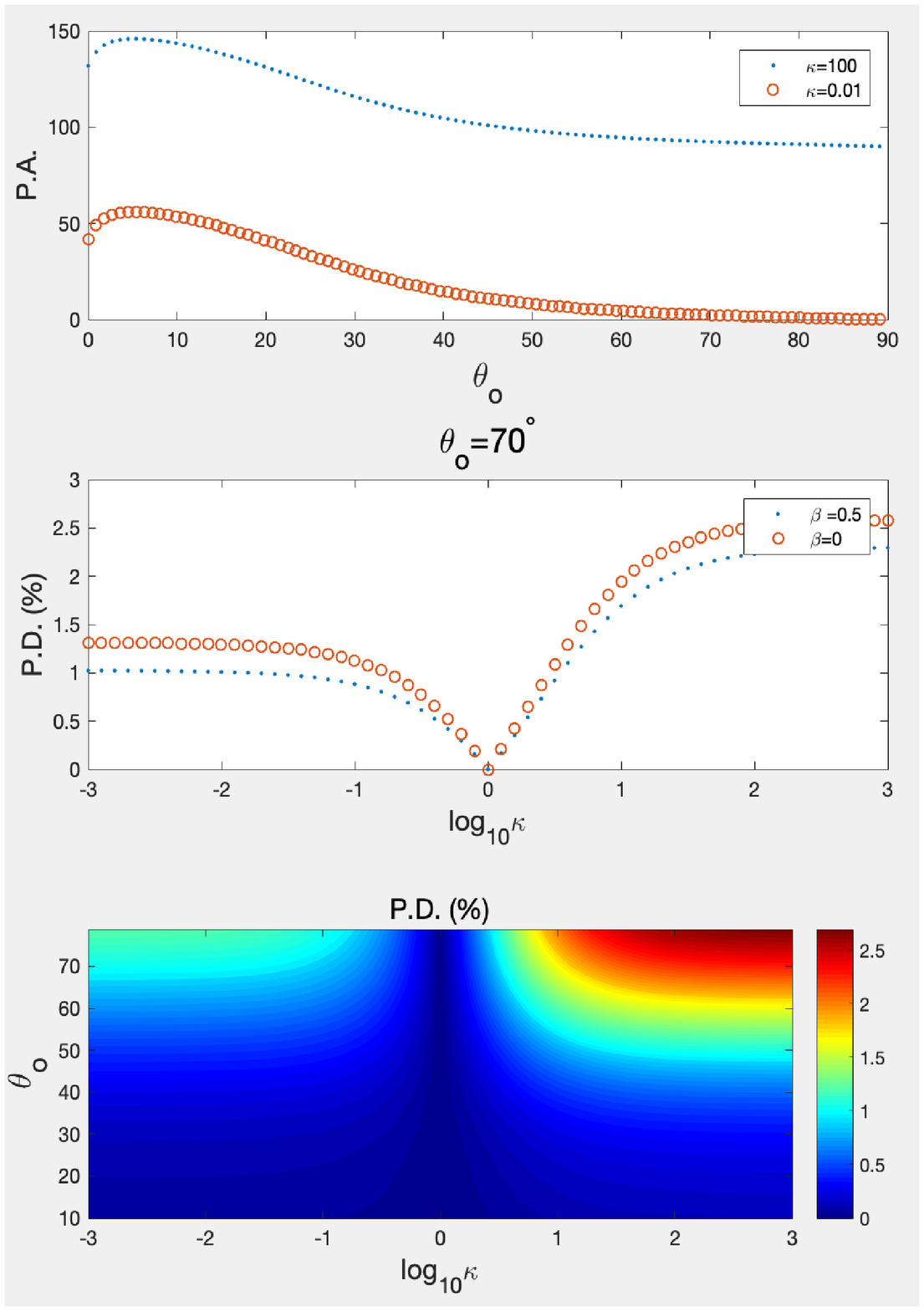}
\caption{Characteristics for orbital averaged P.D. and P.A. with the tangled magnetic field structure.
Top panel: Dependency of P.A. on the inclination angle $\theta_{o}$ for $\kappa=100$ (dotted symbol) and 0.01 (open circle). The results are for $\beta=0.5$ and $\eta=0.01$.
Middle panel: Dependency of P.D. (red color) on the anisotropy parameter $\kappa$ for $\beta=0.5$ (dotted symbol) and 0 (open circle). The results are for $\theta_o=70^{\circ}$ and $\eta=0.01$.
Bottom panel: Contour maps of averaged P.D. on $\theta_o-\kappa$ plane with $\beta=0.5$ .}
\label{syks8}
\end{figure}

Figures~\ref{syks8} show the polarization characteristics with the tangle magnetic field and summarize the dependency on the inclination angle and the anisotropy parameter $\kappa$ in the KS62 model.
As we can see in the top panel of Figure~\ref{syks8}, the dependency of the P.A. for $\kappa>1$ is similar to that of toroidal and poloidal cases (Figure~\ref{sytp}). For $ \kappa<1$, the predicted P.A. is changed by 90 degree from that of $\kappa>1$.

The middle panel of Figure~\ref{syks8} shows the P.D. as a function of the anisotropy parameter $\kappa$ with flow speeds of $\beta=0$ (circle symbol) and 0.5 (dotted symbol). The bottom panel of Figure~\ref{syks8} shows the contour maps of P.D. on $\theta_o-\kappa$ plane. We can see that the averaged P.D. is less than 10~\%.

\section{Summary and discussion}
In this paper, we studied the linear polarization of the emission from gamma-ray binaries based on the pulsar scenario. We assumed that the X-ray emission from the gamma-ray binaries is produced by the synchrotron emission from the shocked pulsar wind electrons/positrons and discussed the polarization properties with three kinds of magnetic field structures: (i) toroidal field, (ii) poloidal field and (iii) tangled field modeled by KS62, which are axisymmetric around the shock-cone axis. Because of the axisymmetric structure of the shock region, the total emission at each orbital phase is linearly polarized in the direction of or perpendicular to the shock-cone axis projected on the sky. For the toroidal field structure, we found that the emission integrated over the shock cone is always oriented along the shock-cone axis on the sky, and the P.A. smoothly changes by $360^{\circ}$ over the orbital phase.
For the poloidal field and tangled field, on the other hand, the direction of the polarization along or perpendicular to the shock-cone axis depends on the system parameters (e.g., flow speed, momentum ratio $\eta$, etc.) and the orbital phase.  At the phase where the P.A. changes by 90$^{\circ}$, we found for the poloidal/tangled fields that the emission can be unpolarized due to the Doppler boosting effect. The P.D. of the toroidal field can approach the maximum value $\Pi_0$ of the synchrotron radiation at the orbital phase, where the shock-cone axis directs $\sim 90^{\circ}$ from the direction of the observer (Figure~\ref{syto}). For the poloidal field, the maximum P.D. is smaller than that of the toroidal field, but it can still reach $\sim 20-30$\% at $~\phi=0.3$ and $0.7$ (Figure~\ref{sypl}). For the tangled field of the KS62 model, the maximum P.D. over the orbital phase depends on the anisotropy parameter $\kappa$, but it is at most $\sim 20$~\% (Figures~\ref{syks1} and~\ref{syks2}). Finally, we applied our model to the gamma-ray binary LS~5039 and predicted the orbital evolution of the polarization characteristics predicted by the different magnetic field structures (Figure~\ref{symd}).

We demonstrated that if the pulsar scenario of LS~5039 is true, a measurement of the polarization can be used to diagnose the magnetic field structure of the shock region. There are several ongoing projects for the measurement of the polarization in X-ray and soft-gamma-ray bands \citep{2016SPIE.9905E..1QZ,2016SPIE.9905E..17W,2019BAAS...51g.245M}. For example, the Imaging X-ray Polarimetry Explorer (IXPE) mission \citep{2019mbhe.confE..68F} can perform spatially resolved spectroscopy in 2-8 keV energy bands. For LS~5039, the observed flux in the 2-8~keV energy bands is on the order of $10^{-11}~{\rm erg~cm^{-2}~s^{-1}}$. According to the sensitivity of the IXPE\footnote{https://directory.eoportal.org/web/eoportal/satellite-missions/i/ixpe}, we can see that several days or several ten-day observations are required to measure the linear polarization with P.D. $\sim$10~\% and 3~\%, respectively. With the orbital parameters fitting the X-ray light curve, our model predicts that the P.D. for the toroidal magnetic field structure can be larger than 10~\% for most of the orbital phase (Figure~\ref{symd}). Moreover, the P.D. averaged over the orbital phase is still larger than 10~\%. We therefore expect that future observations can determine the polarization properties of LS~5039 with a reasonable exposure time if the magnetic field structure of the shock region is described by the toroidal field. For poloidal/tangled magnetic field structure, the P.D. is less than 10~\% for most of the orbital phase, and the P.D. averaged over the orbital phase is smaller than 10\%. Hence, if the magnetic field of the shock region is dominated by the poloidal field or tangled field, no significant detection of the linear polarization will be expected by an observation of several days, except for the orbital phase around of the peak of the P.D.

In this paper, we mainly focused on the study of the polarization characteristics of the X-ray emission of a gamma-ray binary system with simple dynamics of the shocked pulsar wind.
In a subsequent study, we will extend our study with a more appropriate treatment of the dynamics of the shocked pulsar wind (e.g., evolution of the particle distribution under the cooling process) and will predict polarization properties with a more precise comparison with the light curve and spectrum in the multiwavelength bands. 

We thank the reviewer for his/her useful comments and suggestions. We also thank to Prof. K.S. Cheng and Dr A.M. Chen for useful discussion.
H.X.X. and J.T. are supported by the National Key Research and Development Program of China (grant No. 2020YFC2201400) and the National Natural Science Foundation of China (grant No. U1838102).

\bibliographystyle{aasjournal}
\bibliography{b1}


\end{document}